%% file: VBSF.tex
\documentclass[journal]{IEEEtran}

\usepackage{graphicx}
\usepackage{amssymb}
\usepackage{support-caption}
\usepackage{subcaption}

\usepackage{multicol}
\usepackage[cmex10]{amsmath}
\usepackage{amsfonts,mathtools}
\usepackage{url}
\usepackage{array}
\usepackage{url}
\usepackage{bm}
\usepackage{graphicx}
\usepackage{mathrsfs}
\usepackage{amsthm} 
\usepackage{bm}
\usepackage{tikz}
\usepackage{cite}
\usepackage{color}

\usepackage[linesnumbered,ruled]{algorithm2e}
\providecommand{\abs}[1]{\lvert#1\rvert}												
\newcommand{\Ex}[1]{\ensuremath{\mathbb{E}[#1]}} 		  
\newcommand{\ind}{1\hspace{-1.6mm}1}														

\renewcommand{\vec}[1]{\text{vec}\left(#1\right)}

\providecommand{\Diag}[1]{\text{Diag}\left(#1\right)}
\providecommand{\tr}[1]{\text{tr}\left(#1\right)}

\providecommand{\norm}[1]{\left \| #1 \right \|}

\def \b {\mathbf{b}}
\def \a {\mathbf{a}}
\def \A {\mathbf{A}}
\def \B {\mathbf{B}}
\def \D {{\mathcal{D}}}

\def \Y {\mathbf{Y}}
\def \y {\mathbf{y}}
\def \x {\mathbf{x}}
\def \I {\mathbf{I}}
\def \J {\mathbf{J}}
\def \L {\mathbf{L}}
\def \E {\mathbf{E}}
\def \e {\mathbf{e}}
\def \D {\mathbf{D}}
\def \j {\mathbf{j}}
\def \Rn {\mathbb{R}}
\def \Gam {\boldsymbol{\Gamma}}
\def \Lam {\boldsymbol{\Lambda}}
\def \gam {\boldsymbol{\gamma}}
\def \N {\mathcal{N}}

\def \H {\mathcal{H}}

\def \mub {\boldsymbol{\mu}}
\def \bt {\boldsymbol{\beta}}
\def \up {\boldsymbol{\upsilon}}
\def \al {\boldsymbol{\alpha}}
\def \Sig {\boldsymbol{\Sigma}}
\def \Xib {\boldsymbol{\Xi}}
\def \Pb {\boldsymbol{\Psi}}
\def \v {\mathbf{v}}
\def \dott {\boldsymbol{\cdot}}

\author{Charul $^{1}$, \emph{Student Member}, \emph{IEEE}, Uttkarsha Bhatt$^{2}$ , Pravesh Biyani$^{1}$, \emph{Member}, \emph{IEEE} and Ketan Rajawat$^{3}$, \emph{Member}, \emph{IEEE}
\thanks{$^{1}$The authors are with the Department of Electronics and Communication Engineering, Indraprastha Institute of Information Technology, Delhi,
India.}%
\thanks{$^{3}$K. Rajawat is with the Department of Electrical Engineering, Indian Institute
of Technology, Kanpur, UP, India.}%
}

\begin{document}

	\title{Online Variational Bayesian Subspace Filtering with Applications}

\maketitle	
\input{abs_int}

\input{VBPCF_model.tex}
\input{result_conc_revised_pb_edit}

	\bibliographystyle{IEEEtran}
	\bibliography{pbibfile}	

\end{document}

%% file: abs_int.tex
\begin{abstract}

    Matrix completion and robust principal component analysis have been
    widely used for the recovery of data suffering from missing entries or
    outliers. In many real-world applications however, the data is also
    time-varying, and the naive approach of per-snapshot recovery is both
    expensive and sub-optimal. This paper develops generative Bayesian models that fit sequential multivariate measurements arising from a low-dimensional time-varying subspace. A variational Bayesian subspace filtering approach is proposed that learns the underlying subspace and its state-transition matrix.
    Different from the plethora of deterministic counterparts, the proposed
    approach utilizes  automatic relevance determination priors that obviate
    the need to tune key parameters such as rank and noise power. We also
    propose a forward-backward algorithm that allows the updates to be
    carried out at low complexity. Extensive tests over traffic and
    electricity data demonstrate the superior imputation, outlier rejection,
    and temporal prediction prowess of the proposed algorithm over the
    state-of-the-art matrix/tensor completion algorithms.
\end{abstract}

\section{Introduction}

Sensor measurements are often incomplete, noisy, and replete with outliers arising due to malfunctions or intermittent errors. Imputation of the missing entries and removal/segregation of the outliers is a critical first step that must be carried out prior to any data analytics. Examples of applications that benefit from such a pre-processing step include estimation/prediction of city-wide road traffic, regional air quality, electricity consumption in power distribution networks and foreground-background separation in videos. For most of these applications, the measurements can be arranged in form of a matrix, some of whose entries may be missing or contaminated with outliers. Pertinent approaches model the measurements as arising from a low-dimensional subspace whose recovery allows us to reject the noise and outliers, and impute the missing entries \cite{candes2009exact,balzano2010online,babacan-11,giampouras2017online,candes2011robust,ding2011bayesian}.
\par Many real-world applications, including the aforementioned ones, involve time-varying data that arrives in a sequential manner and must be processed as such. As a result, the data matrices arising in such applications comprise of low-dimensional subspaces that evolve over time. While the classical matrix completion or robust principal component analysis (RPCA) approaches are still applicable to each snapshot of the data, the performance can generally be improved by exploiting the temporal correlations present in the measurements \cite{liu2013tensor,balzano2010online,electrictydata,grasta,roseta}. State-of-the-art approaches for processing time-varying subspaces can mostly be classified into approaches based on tensor completion \cite{liu2013tensor} and regularized matrix completion \cite{NIPS2016_6160}. A common feature of these techniques is their static perspective and the resulting focus on batch processing. In contrast however, the data streaming from the sensors may be inherently dynamic, arising from subspaces that evolve over time. Theoretical guarantees for the dynamic setting have been studied in \cite{XuDevenportNIPS}.
Different from these approaches and closer to the classical time-series modeling, an online forecasting matrix completion approach was proposed in \cite{electrictydata} where the underlying subspace was assumed to follow a linear state-space model and must be learned in an online fashion. Approaches based on matrix completion often involve a number of tuning parameters that must be correctly set in order to avoid over-fitting. However determining these parameters via cross-validation is quite challenging with time-series data, especially in the online setting \cite{electrictydata}. Alternatively, probabilistic learning algorithms have been proposed for the static matrix completion, and are generally free of tuning parameters. Such approaches entail constructing generative models that are not only capable of modeling the data but are also simple enough to allow low-complexity updates.

\par
This work considers the first low-rank robust subspace filtering approach for online matrix imputation and prediction. Different from the existing matrix and tensor completion formulations, we consider low-rank matrices whose underlying subspace evolves according to a state-space model. As incomplete columns of the data matrix arrive sequentially over time, the low rank components as well as the state-space model are learned in an online fashion using the variational Bayes formalism. In particular, component distributions are chosen to allow automatic relevance determination (ARD) and unlike the matrix or tensor completion works, the algorithm parameters such as rank, noise powers, and state noise powers need not be specified or tuned. A low-complexity forward-backward algorithm is also proposed that allows the updates to be carried out efficiently. Enhancements to the proposed algorithm, capable of learning time-varying state-transition matrices, operating with a fixed lag, and robust to outliers, are also detailed.  Our approach is general and we demonstrate its efficacy on various settings.  In particular, we discuss the traffic estimation problem in detail and show that the variational Bayesian approach can be used to impute road traffic densities in an online fashion and from only a few observations. As the proposed models are generative, the resulting traffic density predictions can also be used to obtain accurate expected time-of-arrival (ETA) estimates. Additionally, the applicability of the proposed algorithm on the electricity load estimation and prediction problem is also shown. The superior performance of our algorithm vis-a-vis other state of the art subspace tracking and online matrix factorisation algorithms may be attributed to the proposed state space model as well as the flexibility in the data modeling provided by the variational Bayesian approach. In summary, the contributions of the present work are as follows:
\begin{enumerate}
	\item We present the variational Bayesian subspace filtering (VBSF) algorithm and demonstrate its ability to perform data modeling, imputation and temporal prediction in an online setting wherein the key algorithmic parameters are automatically tuned.
	\item Robust version of the VBSF algorithm is also proposed for outlier removal and data cleansing.
	\item Finally, we report a comprehensive comparison of our algorithm with various relevant (offline) matrix completion as well as online subspace estimation and tracking techniques, e.g, GROUSE \cite{balzano2010online}, Low Rank Tensor Completion (LRTC) \cite{liu2013tensor},  GRASTA \cite{grasta}, ROSETA \cite{roseta}, OP-RPCA \cite{oprca} and Online Forecasting Matrix Factorisation (OFMF) \cite{paperarnew} over real-world traffic speed data as well as the electricity load data. 
\end{enumerate}
\subsection {Related work}
Variational Bayesian approaches for matrix completion and robust principal component analysis are well known \cite{babacan-11,Parker-14, Parker-14-2,Wipf-16,yang2018fast,asif2016matrix,luttinen2013fast,giampouras2017online,ma2015variational}. One of the first works considered the measured matrix to be expressible as a product of low-rank matrices, associated with appropriate ARD priors \cite{babacan-11} while faster algorithms for similar settings were proposed in \cite{Parker-14, Parker-14-2}. More recently, other approaches towards modeling the measured matrices have also been proposed \cite{Wipf-16}, \cite {yang2018fast}. Moreover, variational Bayesian approaches have also been applied to road traffic estimation; see e.g. \cite{asif2016matrix}. However, these approaches do not explicitly model the evolution of the underlying subspace. Likewise, none of the existing variational Bayesian approaches for low rank matrix completion model the evolution of the subspace \cite{babacan-11,ma2015variational,yang2018fast}.  In contrast to these, the state-space modeling in our work is inspired from \cite{luttinen2013fast}, where the low-complexity updates were first proposed in the context of linear dynamical models. The VBSF algorithm in the current work extends and generalizes that in \cite{luttinen2013fast} to incorporate low-rank structure and outliers.

On a related note, temporal evolution of the additive noise is modeled in \cite{giampouras2017online} using a forgetting factor. Different from \cite{giampouras2017online} however, we use a state-space model to capture the evolution of the underlying subspace. An online Bayesian matrix factorization model is also proposed in \cite{oprca} wherein the time-stamps are directly incorporated as features. In contrast, the present model is more specific and suited to a slowly time-varying system.

\par
Several non-Bayesian algorithms have been proposed to address the online subspace estimation problem from incomplete observations\cite{balzano2010online,grasta,oprca,roseta}. 
GROUSE \cite{balzano2010online} is one of the early approaches that uses an update on the Grassmannian manifold to estimate the subspace. 
The robust variant of GROUSE, namely GRASTA , handles outliers by by incorporating the $l_1$ norm cost function\cite{grasta}. OP-RPCA\cite{oprca} is a robust subspace estimation technique that uses alternating minimization to compute the outliers and the underlying subspace. A number of online subspace tracking algorithms, such as ROSETA \cite{roseta}, have since been proposed. The proposed approach is compared with some of these algorithms in Sec. \ref{results}. 
\subsection{Applications:}
\subsubsection{Traffic Estimation and Prediction}
Traffic estimation and prediction are the central components of any urban traffic congestion management system \cite{survey_paper}. 
With the advent of smartphones, public transportation services as well as private on-demand transportation companies are increasingly relying on the availability of real-time traffic maps for resource allocation and logistics \cite{res_allocation}. 
Such providers rely on probe vehicles --- GPS enabled and possibly crowd-sourced agents that upload speed measurements and corresponding location tags at sporadic times. Since traffic densities are inferred from speed measurements, they are often ridden with outliers, e.g., corresponding to random velocity changes unrelated to traffic. The traffic estimation problem entails estimating traffic densities at locations and times where no measurements are available. Finally, prediction of traffic in the near future is necessary to calculate ETA, fastest route, and other related quality of service metrics for road users. The future traffic prediction problem becomes particularly challenging in regions with diverse modes of transport, such as in India, where ETA calculations must account for the multimodal nature of traffic \cite{mohan2013moving,goel2016access}. For instance the ETA calculations for buses should not only use traffic data meant for cars.A class of pertinent approaches have sought to visualize the traffic data as an incomplete matrix or tensor, and exploited this correlation to fill-in the missing entries \cite{qu2009ppca,qu2008bpca,tan2016short,asif2016matrix}. Complementary to these approaches, time-series modeling focuses on learning the temporal dynamics of traffic and generate predictions in an online manner \cite{guo2014adaptive}. While recent variants have incorporated spatial correlations as well, these techniques are generally unable to handle missing data or outliers. Finally, \cite{paperarnew} presents the online forecasting matrix factorisation algorithm on the time series data that also handles the missing data scenario.
\subsubsection{Electricity Load Estimation and Prediction}
Similar to the traffic data, the electricity load data also exhibits the spatial and temporal structure that can be exploited to impute the missing data while simultaneously removing the noisy outliers. Due to the environmental disturbance, communication error or sensor fault, it is inevitable that load data may be lost during the collection process \cite{zhang2018short}.
This paper is organized as follows. Sec. \ref{vbsf} presents the online variational Bayesian subspace filtering method for traffic estimation and prediction. Sec. III presents the  online robust variational Bayesian subspace filtering method for traffic estimation and prediction in case of outliers. Results and findings for traffic prediction and electricity load prediction are discussed in Sec. \ref{results} followed by conclusion in Sec. \ref{conclusion}.

%% file: VBPCF_model.tex
\textbf{Notation:} Scalars are denoted by letters in regular font, while vectors (matrices) are denoted by bold face (capital) letters. For a matrix $\A$, its transpose and trace are denoted by $\A^T$ and $\tr{\A}$, respectively. The $(i,j)$-th element of a matrix $\A$ is denoted by $a_{ij}$, the $i$-th column by $\a_{i}$ or $[\A]_{\dott i}$, and the $i$-th row by $\a_{i\dott}^T$ or $[\A]_{i\dott}^T$.  The all-one vector of size $n\times 1$ is represented by $\mathbf{1}_n$, while $\I_n$ denotes identity matrix of size $n \times n$. The Frobenius norm for a matrix $\A$ and the Euclidean norm for a vector $\a$ are denoted by $\norm{\A}$ and $\norm{\a}$, respectively. The multivariate Gaussian probability density function (pdf) with mean vector $\mub$ and covariance matrix $\Sig$ evaluated at $\x \in \Rn^n$ is denoted by $\N(\x \mid \mub,\Sig)$. Likewise, Ga$(x, a, b)$ denotes the Gamma pdf with parameters $a_x$ and $b_x$ evaluated at $x \in \Rn_{+}$. The expectation operator is symbolized by $\mathbb{E}$ while the pdf is generically denoted by $p(\cdot)$. Given data $\D$, the posterior mean is given by $\hat{\x} := \Ex{\x \mid \D}$.

\section{Variational Bayesian Subspace Filtering}
\label{vbsf}
We consider a scenario where the data with the missing entries is arriving in a sequential manner. The data can be considered  in the form of the matrix $\Y \in \Rn^{m \times t}$, where $t$ denotes the number of time instances over which measurements are made and $m$ denotes the number of rows of the matrix $\Y$. More generally, $\Y$ is an incomplete and growing matrix whose columns  arrive sequentially over time. Specifically, for each column $\y_\tau$ with $1\leq \tau \leq t$, only entries from the index set $\Omega_\tau\subset \{1, \ldots, m\}$ are observed. The algorithms developed here will seek to achieve the following two goals:
\begin{itemize}
	\item \emph{imputation} which yields $\{\hat{y}_{i\tau}\}_{i\notin\Omega_\tau}$ for $1\leq \tau \leq t$, and
	\item \emph{prediction} which yields $\{\hat{\y}_{t+\tau}\}_{\tau = 1}^{T_p}$ where $T_p$ is the prediction horizon. 
\end{itemize} 
The next subsection develops a variational Bayesian algorithm for achieving the aforementioned goals. 	

\begin{figure}[ht!]
	\centering
	\includegraphics[width=0.9\linewidth]{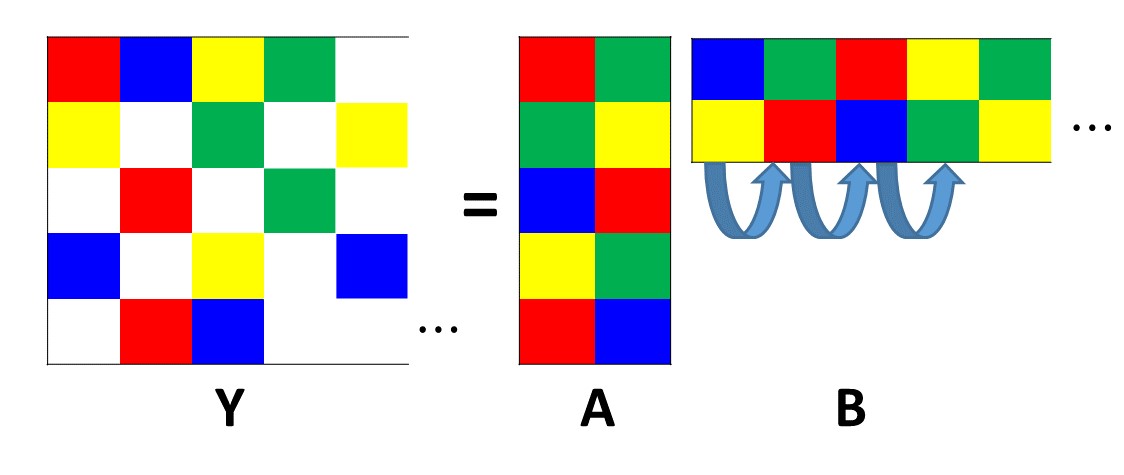}
	\caption{Online Variational Bayesian Filtering}
	\label{fig:matfac}
\end{figure}
\subsection{Hierarchical Bayesian Model}
We begin with detailing a generative model for the matrix $\Y$. The proposed model will not only capture the rank deficient nature of $\Y$ \cite{babacan2012sparse} but also the temporal correlation between successive columns of $\Y$ \cite{asif2013low}. Recall that the standard low-rank parametrization of the full  matrix $\Y$ takes the form $\Y = \A\B$ where $\A \in \Rn^{m \times r}$ and  $\B \in \Rn^{r \times t}$. Classical non-negative matrix completion approaches seek to obtain such a factorization. In such algorithms, the choice of $r$ is critical to avoiding underfitting or overfitting. \par
Within the Bayesian setting however, the measurements are modeled as arising from a distribution with unknown hyper-parameters, while various components or parameters are assigned different prior distributions. The Bayesian framework allows the use of ARD, wherein associating appropriate priors to the model parameters leads to pruning of the redundant features \cite{babacan2012sparse}. This work uses pdfs from the exponential family that allow for tractable forms of the posterior pdf but are also flexible enough to adequately model the data. \par 
Specifically, the entries of $\Y$ are generated as
\begin{align}
p(y_{i\tau} \mid \a_{i\dott}, \b_\tau, \beta) &= \N(y_{i\tau} \mid \b_\tau^T\a_{i\dott}, \beta^{-1}) 
&  i \in \Omega_{\tau}
\end{align}
for all $\tau \geq 1$, where $\A \in \Rn^{m \times r}$, $\B \in \Rn^{r \times t}$, and $\beta \in \Rn_{++}$ are the (hidden) problem parameters. Unlike the deterministic setting however, the rank hyper-parameter $r$ is not critical to the imputation or prediction accuracy, but is only required to chosen according to computational considerations. The temporal evolution of the entries of $\Y$ is modeled by making the columns of $\B$ adhere to the following first order autoregressive model: 
\begin{align}\label{ss}
p(\b_\tau \mid \J, \b_{\tau-1}) &= \N(\b_{\tau} \mid \J\b_{\tau-1}, \I_r) & 2\leq \tau\leq t
\end{align} 
for $\tau \geq 2$, where $\J \in \Rn^{r \times r}$ is again a problem parameter. 
Here, $\J$ captures the temporal structure of the underlying subspace, and is learned from the data itself. The scaling ambiguity present in matrix factorization allows the transition matrix $\J$ to capture both slow and fast variations in $\b_\tau$ without the need to explicitly model the state noise variance. 
It follows from \eqref{ss} that the conditional pdf of $\b_\tau$ given $\J$ is given by 
\begin{align}
p(\B \mid \J) = \N(\b_1; \mub_1, \Lam_1 ) \prod_{\tau = 2}^t \N(\b_\tau \mid \J\b_{\tau-1}, \I_r).
\end{align} 
Observe that the model complexity depends on the rank $r$, which is also the number of columns in $\A$ and $\J$. In order to ensure the value of $r$ is learned in a data-driven fashion, the columns of $\A$ and $\J$ are assigned multivariate Gaussian priors with column-specific precisions, i.e., 
\begin{align}
p(\A \mid \gam) &= \prod_{i=1}^r \N(\a_i \mid 0, \gamma_i^{-1}\I_m) \label{paa}\\
p(\J \mid \up) &= \prod_{i=1}^r \N(\j_i \mid 0, \upsilon_i^{-1}\I_r) \label{pja}
\end{align}
where the precisions $\gam$ and $\up$ are problem parameters. It can be seen that if any of $\gamma_i$ or $\upsilon_i$ are large, the corresponding columns will be close to zero and consequently irrelevant. Indeed, the priors in \eqref{paa}-\eqref{pja} aid in automatic relevance determination since the subsequent optimization process may drive some of the precisions to infinity, yielding a low-rank factorization. 

\begin{figure}[ht!]
	\centering
	\includegraphics[width=0.9\linewidth]{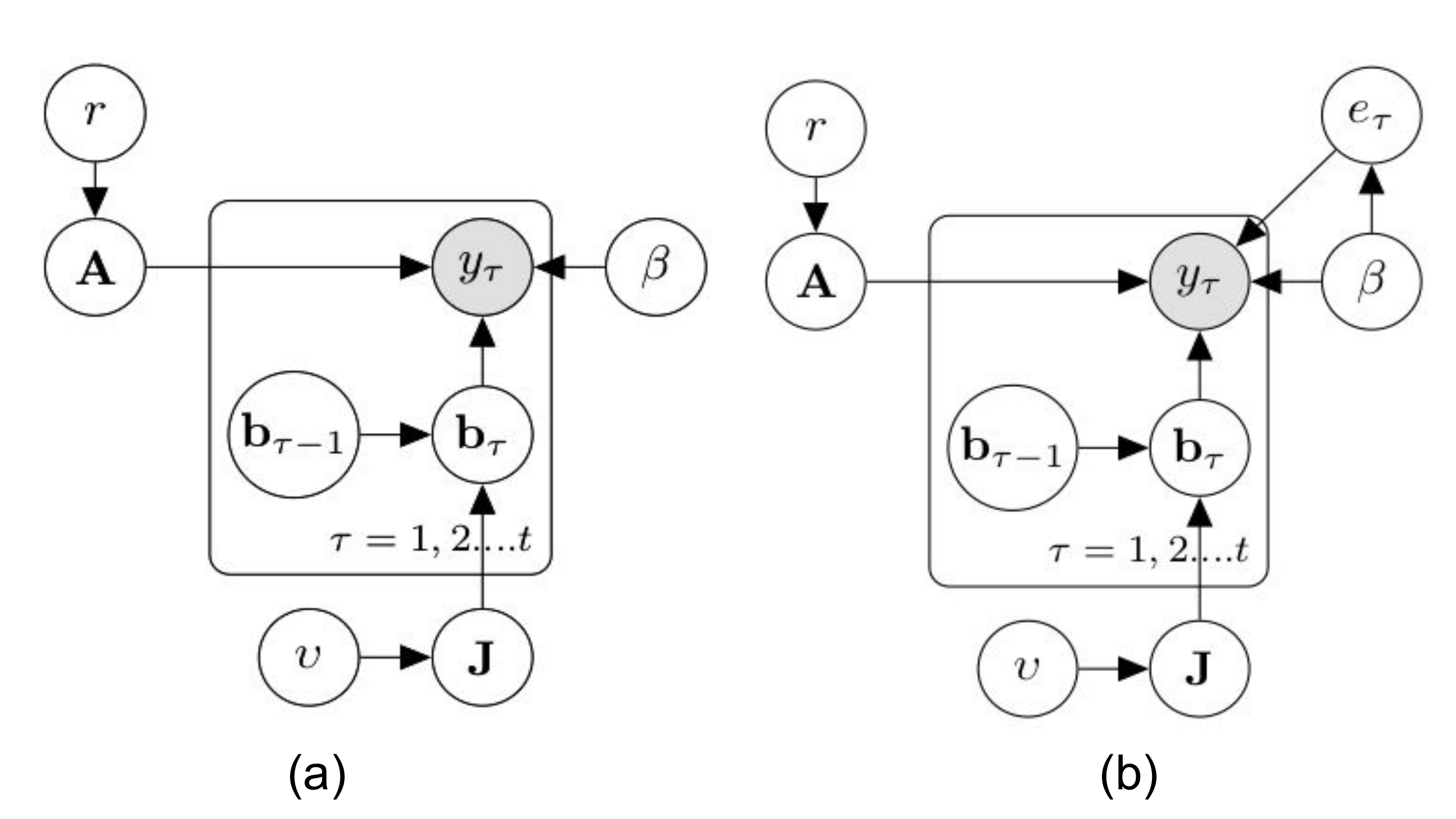}
	
	\caption{(a) Hierarchical Bayesian Model for Matrix Completion (b) Robust Hierarchical Bayesian Model for Matrix Completion }
	\label{fig:mc_algo}
\end{figure}

Finally, the three precision variables are selected to have have non-informative Jeffrey's priors
\begin{align}
p(\beta) &= \frac{1}{\beta}, & p(\gamma_i) &= \frac{1}{\gamma_i}, & p(\upsilon_i) &= \frac{1}{\upsilon_i}
\end{align}
for $1\leq i \leq r$. Let $\y_{\Omega}$ denote the collection of measurements $\{y_{i\tau}\}_{i\in\Omega_\tau, \tau = 1}^t$. Collecting the hidden variables into $\H := \{\A, \B, \J, \beta, \gam, \up\}$, the joint distribution of $\{\y_\Omega, \H\}$ can be written as
\begin{align}
p(\y_\Omega,\H) &= p(\y_\Omega | \A, \B, \beta)p(\A | \gam)p(\B | \J) p(\J | \up)p(\beta)p(\up)p(\gam) \nonumber \\
&=\prod_{\tau=1}^t\prod_{i\in\Omega_\tau} \N(y_{i\tau} \mid \b_\tau^T\a_{i\dott}, \beta^{-1}) \nonumber\\
&\times \prod_{i=1}^r \left[\N(\a_i \mid 0,\gamma_i^{-1}\I_m) \N(\j_i \mid 0, \upsilon_i^{-1}\I_r)\right]  \nonumber \\
&\hspace{-1cm}\times\N(\b_1; \mub_1, \Lam_1 ) \prod_{\tau = 2}^t \N(\b_\tau \mid \J\b_{\tau-1}, \I_r) \frac{1}{\beta}\prod_{i=1}^r \frac{1}{\gamma_i\upsilon_i}
\end{align} 
The full hierarchical Bayesian model adopted here is summarized in Fig. \ref{fig:mc_algo}(a). 

\subsection{Variational Bayesian Inference}
Having specified the generative model for the data, the goal is to determine the posterior distribution $p(\H|\y_\Omega)$, which would yield the corresponding point estimates and can be used for imputation and prediction tasks. However, exact full Bayesian inference is well-known to be intractable. Instead, we utilize the mean-field approximation, wherein the posterior distribution factorizes as:

\begin{align}\label{mf}
p(\H \mid \y_\Omega) \approx q(\H) = q_{\A}(\A)q_{\B}(\B)q_{\J}(\J)q_{\up}(\up)q_{\beta}(\bt)q_{\gam}(\gam).
\end{align}

In other words, the posterior is now restricted to a family of distributions that adhere to \eqref{mf}. The factors $q_\A$, $q_\B$, $q_\J$, $q_\upsilon$, $q_\beta$, and $q_\gamma$ can be determined by minimizing the Kullback--Leibler divergence of $p(\H|\y_\Omega)$ from $q(\H)$, usually via an alternating minimization approach \cite{bishop2006pattern}.  
Indeed, thanks to the choice of conjugate priors for the parameters, it can be shown that the individual factors in \eqref{mf} take the following forms \cite{luttinen2013fast}:

\begin{subequations}\label{qs}
	\begin{align}
	q_{\B}(\B) &= \N(\vec{\B} \mid \mub^{\B}, \Xib^{\B}) \\
	q_{\a_{i\dott}} &= \N({\a_{i\dott}} \mid \mub_i^{\A}, \Xib_i^{\A}) \\
	q_{\j_{i\dott}} &= \N({\j_{i\dott}} \mid \mub_i^{\J}, \Xib_i^{\J}) \\
	q_{\beta}(\beta) &= \text{Ga}(\beta; a^\beta, b^\beta) \\
	q_{\gamma_i}(\gamma_i) &= \text{Ga}(\gamma_i; a_i^\gamma, b_i^\gamma) \\
	q_{\upsilon_i}(\upsilon_i) &= \text{Ga}(\upsilon_i; a_i^\upsilon, b_i^\upsilon) 
	\end{align}
\end{subequations}
where, $\mub^\B \in \Rn^{rt}$,  $\Xib^{\B} \in \Rn^{rt \times rt}$, $\mub^\A_i \in \Rn^r$, $\Xib^\A_i \in \Rn^{r\times r}$, $\mub^\J_i \in \Rn^r$, $\Xib^\J_i \in \Rn^{r\times r}$, and $a^\beta$, $b^\beta$, $a^{\gamma}_i$, $b^{\gamma}_i$, $a^\upsilon_i$, $b^\upsilon_i \in \Rn_{++}$. Consequently, each iteration of alternating optimization simply involves updating the variables $\{\mub^\B, \Xib^\B, \{\mub^\A_i\}, \{\Xib^\A_i\}, \{\mub^\J_i\}, \{\Xib^\J_i\}$, $a^\beta, b^\beta$, $\{a^{\gamma}_i\}, \{b^{\gamma}_i\}, \{a^\upsilon_i\}, \{b^\upsilon_i\}\}$ in a cyclic manner. 

In the present case, not all variables need to be updated explicitly and the updates may be written in a compact form. 
Let us denote $\omega_\tau:=\abs{\Omega_\tau}$ and let $\omega:=\sum_\tau \omega_\tau$ be the total number of observations made. Then, the updates for hyperparameters $\{\up,\gam\}$ take the following form 
\begin{subequations}\label{upga}
	\begin{align}
	\hat{\upsilon}_i &= \frac{m}{\sum_{k=1}^m\left([\mub^\J_k]^2_i + [\Xib^\J_k]_{ii}\right)} \label{10a} \\
	\hat{\gamma}_i &= \frac{m}{\sum_{k=1}^m\left([\mub^\A_k]^2_i + [\Sig^\A_k]_{ii}\right)}.\label{10b} 
	\end{align}
\end{subequations}
Subsequently, let $\hat{\up}$ and $\hat{\gam}$ be the vectors that collect $\{\hat{\upsilon}_i\}$ and $\{\hat{\gamma}_i\}$, respectively. Since $\b_{\tau}$ denotes the $\tau$-th column of $\B^T$, its posterior distribution may be written as $q_{\b_\tau}(\b_\tau) = \mathcal{N}(\b_\tau \mid \mub^\B_\tau, \Xib^\B_\tau)$, where $\mub^\B_\tau$ and $\Xib^\B_\tau$ comprise of the corresponding elements of $\mub^\B$ and $\Xib^\B$, respectively. Also define the posterior covariance matrices
\begin{align} \label{sigma_up1}
\Sig^{\B}_{\tau,\iota} &:= \mub^\B_\tau(\mub^\B_\iota)^T + \Xib^{\B}_{\tau,\iota} \\
\label{sigma_up2}
\Sig^\J_i &:= \mub^\J_i(\mub^\J_i)^T + \Xib^\J_i \\
\label{sigma_up3}
\Sig^\A_i &:= \mub^\A_i(\mub^\A_i)^T + \Xib^\A_i.
\end{align}
Therefore, the update for $\hat{\beta}$ becomes
\begin{align}\label{be}
\hat{\beta} = \frac{\omega}{\sum_{\tau=1}^t\sum_{i\in\Omega_\tau} \left[y_{i\tau}^2 - 2y_{i\tau}(\mub^\A_i)^T\mub^\B_\tau + \tr{\Sig^\A_{i}\Sig^{\B}_{\tau,\tau}}\right]}.
\end{align}  

Next, the updates for the factors $\J$ and $\A$ take the following form
\begin{subequations}\label{ja}
	\begin{align}
	\mub^\J_i &= [\Xib^\J_i\Sig^{\B}_{\tau,\tau-1}]_{\dott i} \label{15a}\\
	\Xib^\J_i &= \left(\Diag{\hat{\up}} + \sum_{\tau=1}^{t-1}\Sig^\B_{\tau,\tau-1}\right)^{-1}\label{15b} \\
	\mub^\A_i &= \hat{\beta}\Xib^\A_i\sum_{\tau \in\Omega'_i} \mub^\B_\tau y_{i\tau}\label{15c} \\
	\Xib^\A_i &=  \left(\hat{\gamma}_i\I_{r} + \hat{\beta}\sum_{\tau \in\Omega'_i}\Sig^\B_{\tau,\tau} \right)^{-1}\label{15d}
	\end{align}
\end{subequations}
where $\Omega'_i:=\{\tau \mid i \in \Omega_\tau\}$. Observe from the updates that the rows of $\J$ are independent identically distributed under the mean field approximation. The update for $\mub^\B$ can be written as
\begin{align}
\mub^\B &= \Xib^\B\begin{bmatrix} \hat{\beta}\sum_{i\in\Omega_1}y_{i1}\mub^\A_i  + \Lam_1^{-1}\mub_1\\
\hat{\beta}\sum_{i\in\Omega_2}y_{i2}\mub^\A_i \\
\vdots\\
\hat{\beta}\sum_{i\in\Omega_t}y_{it}\mub^\A_i
\end{bmatrix}\label{mub}.
\end{align}
Finally, $[\Xib^\B]^{-1}$ a block-tridiagonal matrix. Defining $\hat{\J}:=\Ex{\J \mid \y_\Omega}$ as the matrix whose $i$-row is given by $(\mub^\J_{i})^T$, $\Sig^\A_{(\tau)} = \sum_{i\in\Omega_\tau'}\Sig^\A_i$, and $\Sig^\J:=\sum_{i=1}^r \Sig^\J_i$, the updates take the form:
\begin{align}
\left[\Xib^{\B}\right]^{-1} &=  \hat{\beta}\Diag{\Xib^\A_{(1)}, \ldots, \Xib^\A_{(t)}}  + \nonumber\\
&+ \begin{bmatrix} \Lam_1^{-1} & -\hat{\J} & \ldots &0\\
-\hat{\J} & \I_r + \Sig^\J & -\hat{\J}  & \ldots \\
\vdots & \vdots &&\vdots \\
\ldots & 0 & -\hat{\J} &  \I_r 
\end{bmatrix}. \label{xib}
\end{align}
It is remarked that although the $rt \times rt$ matrix $[\Xib^\B]^{-1}$ is block-tridiagonal, the matrix $\Xib^\B$ is dense, and direct inversion would be prohibitively costly. Moreover, the classical Rauch-Tung-Striebel (RTS) smoother cannot be directly applied since evaluating the conditional expectations under $q(\B)$ is difficult and not amenable to the Matrix Inversion Lemma \cite{beal2003variational}. Interestingly, observe that the updates in \eqref{be} and \eqref{ja} depend only on diagonal and super-diagonal blocks of $\Xib^\B$, namely $\Xib^\B_{\tau,\tau}$ and $\Xib^\B_{\tau,\tau-1}$, respectively. The next subsection details a low-complexity algorithm for carrying out the updates for these blocks as well as for $\mub^\B$.
\subsection{Low-complexity updates via LDL-decomposition}
Thanks to the block-tridiagonal structure of $[\Xib^\B]^{-1}$, it is possible to use the LDL decomposition to carry out the updates in an efficient manner. Decomposing $[\Xib^\B]^{-1} = \L\D\L^T$, the key idea is that left multiplication with $\Xib^\B$ is equivalent to left multiplication with $\L^{-T}\D^{-1}\L^{-1}$. Towards this end, we utilize the algorithm from \cite{luttinen2013fast}, that comprises of two phases: the forward pass that carries out the multiplication with $\D^{-1}\L^{-1}$ and the backward pass that implements the multiplication with $\L^{-T}$. Let us define for $2\leq \tau \leq t$, 
\begin{align}
\Pb_\tau &:= \hat{\beta}\sum_{i\in\Omega_\tau}\Sig^\A_{(i)}+ \I_r + \ind_{\tau\neq t}\sum_{i=1}^r \Sig^\J_i	\\
\v_\tau &:= \hat{\beta}\sum_{i\in\Omega_\tau} y_{i\tau}\mub^\A_i.\label{vbt}
\end{align}
The forward pass outputs intermediate variables $\breve{\Xib}^{\B}_{\tau,\tau}$, $\breve{\Xib}^{\B}_{\tau,\tau+1}$, and $\breve{\mub}_\tau$, that are subsequently used in the backward pass. The updates take the following form:
\begin{enumerate}
	\item Initialize $\hat{\Xib}^{\B}_{1,1} = \Lam_1$ and $\hat{\mub}^{\B}_1 = \mub_1 + \hat{\beta}\sum_{i\in\Omega_\tau}y_{i\tau}\Lam_1\mub^\A_i$
	\item For $\tau = 1, \ldots, t-1$
	\begin{subequations} \label{rts}
		\begin{align}
		\hspace{-1cm}\breve{\Xib}^{\B}_{\tau,\tau+1} & = -\hat{\Xib}^{\B}_{\tau,\tau}\hat{\J} \\
		\hspace{-1cm}\breve{\Xib}^{\B}_{\tau+1,\tau+1} &= (\Pb_{\tau+1} - (\breve{\Xib}^{\B}_{\tau,\tau+1})^T\Pb^{\B}_{\tau,\tau+1})^{-1} \\
		\breve{\mub}^{\B}_{\tau+1} &= \breve{\Xib}^{\B}_{\tau+1,\tau+1}(\v_{\tau+1} - (\breve{\Xib}^{\B}_{\tau,\tau+1})^T\breve{\mub}^{\B}_{\tau})
		\end{align}
		\item For $\tau = t-1, \ldots, 1$
		\begin{align}
		\hspace{-1cm}\Xib^{\B}_{\tau,\tau+1} &= - \breve{\Xib}^{\B}_{\tau,\tau+1}\Xib^{\B}_{\tau+1,\tau+1} \\
		\hspace{-1cm}\Xib^{\B}_{\tau,\tau} &= \breve{\Xib}^{\B}_{\tau,\tau} - \hat{\Xib}^{\B}_{\tau,\tau+1}(\Xib^{\B}_{\tau,\tau+1})^T \\
		\hspace{-1cm}\mub^{\B}_{\tau} &= \breve{\mub}^{\B}_{\tau} - \breve{\Xib}^{\B}_{\tau,\tau+1}\mub^{\B}_{\tau+1}
		\end{align}
	\end{subequations}
	\item Output $\{\Xib^{\B}_{\tau,\tau+1}, \Xib^{\B}_{\tau,\tau}, \mub^{\B}_{\tau}\}_{\tau = 2}^t$
\end{enumerate}
Note that while $\Xib^{\B}_{i,j} \neq 0$ for $|i-j| > 1$, these blocks are neither calculated in the forward and backward passes nor required in any of the variational updates.

Finally, the predictive distribution $p(y_{i\tau}\mid \y_{\Omega})$ for $\tau \notin \Omega_i$ or $\tau \geq t+1$ is still not tractable in the present case. Instead, we simply use point estimates for estimating the missing entries. Specifically, for $\tau \notin \Omega_i$, the missing entries are imputed as
\begin{align}
y_{i\tau} = (\mub^\B_\tau)^T\mub^\A_i. \label{pred1}
\end{align}
Likewise for $\tau \geq t+1$, the prediction becomes
\begin{align}
y_{i\tau} = (\hat{\J}^{\tau-t}\mub^\B_t)^T\mub^\A_i.\label{pred2}
\end{align}

It can be seen that as compared to the updates in \eqref{mub}-\eqref{xib} that incur a complexity of $\mathcal{O}(t^3)$, the complexity incurred due to \eqref{rts} is only $\mathcal{O}(t)$. Overall, the different parameters are updated cyclically until convergence for each $t = 1, 2, \ldots$.

\subsection{EM Baysian Subspace Filtering} 
Different from the variational Bayesian framework used here, the EM algorithm treats $\H_h:=\{\A, \B, \J\}$ as hidden variables (with posterior pdf $q_h(\H_h):=q_\B(\B)q_\A(\A)q_\J(\J)$) and uses maximum a posteriori (MAP) estimates for the precision variables $\H_p:=\{\up, \gam, \beta\}$. Consequently, the EM algorithm for Bayesian subspace tracking starts with an initial estimate $\H_p^{(0)}$ and uses the following updates at iteration $\iota \geq 1$,
\begin{itemize}
	\item \textbf{E-step:} evaluate
	\begin{align}
	Q(\H_p,\H_p^{(\iota)}):= \mathbb{E}_{q_h(\H_h)}\left[\log p(\y_\Omega,\H_h,\H_p^{(\iota)})\right]
	\end{align}
	\item \textbf{M-step:} maximize
	\begin{align}
	\H_p^{(\iota+1)} = \arg\max_{\H_p} Q(\H_p,\H_p^{(\iota)})
	\end{align}
\end{itemize}

Interestingly, the updates resulting from the E-step take the same form as those in \eqref{ja} and \eqref{rts}. On the other hand, the updates obtained from solving the M-step take the slightly different form:
\begin{subequations}\label{mstep}
	\begin{align}
	\hat{\upsilon}_i &= \frac{m-2}{\sum_{k=1}^m\left([\mub^\J_k]^2_i + [\Xib^\J_k]_{ii}\right)} \\
	\hat{\gamma}_i &= \frac{m-2}{\sum_{k=1}^m\left([\mub^\A_k]^2_i + [\Sig^\A_k]_{ii}\right) }\\
	\hat{\beta} &= \frac{\omega-2}{\sum_{\tau=1}^t\sum_{i\in\Omega_\tau} \left[y_{i\tau}^2 - 2y_{i\tau}(\mub^\A_i)^T\mub^\B_\tau + \tr{\Sig^\A_{i}\Sig^{\B}_\tau}\right]}.
	\end{align}
\end{subequations}
The slight differences arise due to the difference between the mean and mode of the Gamma distribution. Specifically, for $p(x) =$ Ga$(x | a,b)$, it holds that $\Ex{X} = a/b$ while $\max_x \text{Ga}(x | a,b) = \frac{a-1}{b}$.  \par 

\subsubsection{Remarks on the Convergence of VBSF}
The VB framework used in the present work is a special case of a more general mean field approximation approach. The convergence of the VB algorithm is well-known; see e.g. \cite{tzikas2008variational}, \cite{sato2001online}. Intuitively, the variational approximation renders the evidence lower bound convex in individual factors, and thus amenable to coordinate ascent iterations. Since the lower bound is also differentiable with respect to each factor, the coordinate ascent iterations converge to a stationary point; see \cite{tseng2001convergence} for a more general result. However, convergence to the global optimum is not guaranteed. 

\subsection{Fixed-lag tracking}
Algorithm \ref{alg:ALGO} can be viewed as an offline algorithm that must be run for every $t$. In practical settings, it may be impractical to remember and process the entire history of measurements at each $t$. Moreover, given data at time $t$, estimates may only be required for entries at time  $t-\Delta$ for some $\Delta < h$. 
Towards this end, we consider a sliding window of measurements.
%
Since $\A_{t}$ and $\J_{t}$ may be seen as transition matrices for the latent states and between latent state and observations, we initialize the next sliding-window with inferred approximate distributions on the transition matrices of the current window.
For instance, within the context of traffic density prediction, the inferred approximate distribution for a day may be used as a prior for the coming days. 
 That is, the distributions for $\A, \, \B,$ and $\J$  for a day and sliding window can be initialized with the approximate distributions obtained from the previous month's data. 
%
%
%
\begin{algorithm}
	Initialize $  \gam,\bt,\v$, $sub=1,\, \Omega_\tau, \,\Omega'_i,\Xib^\A,\mub^\A, \Xib^\B,\mub^\B,\Xib^\J_{diag},\mub^\J \Lam_1,\mu_1, $
	
	$\hat{\mathbf{Y}}=\mub^A(\mub^\B)^T$
	
	\While{$Y_{conv}< 10^{-5}$}{
		$\mathbf{Y_{old}}= \hat{\mathbf{Y}}$
		
		$\Gam=diag(\gam)$
		
		\uIf{$sub==1 $}
		{	
			$\text{Update using} \, \eqref{rts}$
			
			$sub=2$
			
			$ \text{Update using} \, \eqref{10a}, \,\eqref{sigma_up1},\,\eqref{15a},\,\eqref{15b}\,\, \forall \,\, 1\leq i \leq r$
			
		}
		
		\ElseIf{$sub==2 $}  
		{
			$\text{Update using} \, \eqref{sigma_up3},\eqref{15c}, \eqref{15d}, \eqref{10b} \, \, \forall \, 1\leq i \leq m$
			
			$sub=1$
		}		
		$\hat{\mathbf{Y}}=\mub^A(\mub^\B)^T$
		
		$\text{Update using } \eqref{be}$
		
		$Y_{conv}=\frac{\norm{\mathbf{Y}-\mathbf{Y_{old}}}_F}{\norm{\mathbf{Y_{old}}}_F}$
	}
	\Return($\hat{\mathbf{Y}},\Xib^\A,\mub^\A, \Xib^\B,\mub^\B,\Xib^\J_{diag},\mub^\J $)
	
	\caption{Variational Bayesian Subspace Filtering}
	\label{alg:ALGO}
\end{algorithm}

\section{Robust Variational Bayesian Subspace Filtering}\label{rvbsf}
In this section we consider the robust version of the variational Bayesian subspace filtering problem in Sec. \ref{vbsf}. Within this context, in addition to the missing entries in $\Y$, some entries of $\Y$ are also contaminated with outliers. Unlike the missing entries however, the location of these outliers is not known. These entries arise due to sensor malfunctions, communication errors, and impulse noise. The robust subspace filtering problem is more difficult as the removal of such outliers entails estimating their magnitudes as well as locations. 

Within the deterministic robust PCA framework, the  matrix is modeled as taking the form $\Y = \A\B + \E$ where $\A \in \Rn^{m \times r}$, $\B \in \Rn^{r \times t}$ are low-rank matrices as before. Additionally, we also need to estimate the sparse outlier matrix $\E \in \Rn^{m \times t}$. As before, both $r$ and the level of sparsity in $\E$ are tuning parameters that must generally be carefully selected. 

Here, we put forth the variational Bayesian subspace filtering algorithm that makes use of ARD priors to prune the redundant features. Consider the measurement matrix $\Y$, whose entries are generated from the following pdf:
\begin{align}
p(y_{i\tau} \mid \a_{i\dott}, \b_\tau, e_{i\tau}, \beta) &= \N(y_{i\tau} \mid \b_\tau^T\a_{i\dott} + e_{i\tau}, \beta^{-1})  & i \in \Omega_\tau
\end{align}
for all $\tau \geq 1$, and apart from the matrices $\A$ and $\B$ defined earlier, we also have $\{e_{i\tau}\}_{\tau =1, i\in \Omega_\tau}^t$ as the additional (hidden) problem parameter that captures the outliers. The generative models for $\A$ and $\B$ are the same as before, i.e.,
\begin{subequations}\label{pbaj}
	\begin{align}
	p(\B \mid \J) &= \N(\b_1; \mub_1, \Lam_1 ) \prod_{\tau = 2}^t \N(\b_\tau \mid \J\b_{\tau-1}, \I_r)\label{pb}\\
	p(\A \mid \gam) &= \prod_{i=1}^r \N(\a_i \mid 0, \gamma_i^{-1}\I) \label{pa}\\
	p(\J \mid \up) &= \prod_{i=1}^r \N(\j_i \mid 0, \upsilon_i^{-1}\I) \label{pj}
	\end{align} 
\end{subequations}
for $\tau \geq 2$, and $\gam$ and $\up$ are problem parameters. Additionally, we also associate an ARD prior to the outliers, i.e.,
\begin{align}
p(e_{i\tau}) &= \N(e_{i\tau} \mid 0, \alpha_{i\tau}^{-1}) & i \in \Omega_\tau
\end{align}
for $1\leq \tau\leq t$, where the precision $\alpha_{i\tau}$ is a hidden variable, that would be driven to infinity whenever $e_{ij}$ is zero. It is remarked that the prior for $e_{i\tau}$ is only specified for the measurements, i.e., for $i \in \Omega_\tau$ and no predictions are made for the outliers. As before, we associate Jeffery's prior to the precisions $\beta$, $\{\gamma_i\}$, $\{\upsilon_i\}$, and $\{\alpha_{i\tau}\}$. 
\begin{align}
p(\beta) &= \frac{1}{\beta}, & p(\gamma_i) &= \frac{1}{\gamma_i}, & p(\upsilon_i) &= \frac{1}{\upsilon_i}, & p(\alpha_{i\tau}) &= \frac{1}{\alpha_{i\tau}}.
\end{align}

Let the vectors $\e \in \Rn^{\omega}$ and $\al \in \Rn^{\omega}$ collect the variables $\{e_{i\tau}\}$ and $\{\alpha_{i\tau}\}$, respectively. Likewise, defining all the hidden variables as $\H := \{\A, \B, \J, \e, \beta, \gam, \up\}$, the joint distribution of $\{\y_\Omega, \H\}$ can be written as
\begin{align}
p(\y_\Omega,\H) &\nonumber\\
&\hspace{-1.5cm}= p(\y_\Omega | \A, \B, \beta)p(\A | \gam)p(\B | \J) p(\J | \up)p(\e | \al)p(\beta)p(\up)p(\gam) \nonumber \\
&\hspace{-1cm}=\prod_{\tau=1}^t\prod_{i\in\Omega_\tau} \N(y_{i\tau} \mid \b_\tau^T\a_{i\dott}, \beta^{-1}) \N(e_{i\tau}\mid 0,\alpha_{i\tau}^{-1})\frac{1}{\alpha_{i\tau}}\nonumber\\
&\times \prod_{i=1}^r \left[\N(\a_i \mid 0,\gamma_i^{-1}\I) \N(\j_i \mid 0, \upsilon_i^{-1}\I)\right]  \nonumber \\
&\hspace{-1cm}\times\N(\b_1; \mub_1, \Lam_1 ) \prod_{\tau = 2}^t \N(\b_\tau \mid \J\b_{\tau-1}, \I) \frac{1}{\beta}\prod_{i=1}^r \frac{1}{\gamma_i\upsilon_i}.
\end{align} 
The full hierarchical Bayesian model adopted here is summarized in figure \ref{fig:mc_algo}(b).

\subsection{Variational Bayesian Inference}\label{vbi2}
Utilizing the mean field approximation, the posterior distribution $p(\H \mid \y_\Omega)$ factorizes as
\begin{align}\label{mf2}
p(\H \mid \y_\Omega) \approx q(\H) &\nonumber\\
&\hspace{-2cm}= q_{\A}(\A)q_{\B}(\B)q_{\J}(\J)q_{\e}(\e)q_{\up}(\up)q_{\beta}(\bt)q_{\gam}(\gam).
\end{align}
where the individual factors take the same forms as in \eqref{qs}, in addition to
\begin{align}
q_{\e}(\e) & = \prod_{\tau = 1}^t \prod_{i\in \Omega_\tau} \N(e_{i\tau} | \mu_e^{i\tau}, \Xi_e^{i\tau}).
\end{align}
As before, the variational inference problem can be solved by updating the variables  $\{\mub^\B, \Xib^\B, \{\mub^\A_i\}, \{\Xib^\A_i\}, \{\mub^\J_i\}, \{\Xib^\J_i\}$, $\{\mu_e^{i\tau}\}, \{\Xi_e^{i\tau}\}, a^\beta, b^\beta$, $\{a^{\gamma}_i\}, \{b^{\gamma}_i\}, \{a^\upsilon_i\}, \{b^\upsilon_i\}\}$ in a cyclic manner. However, a more compact form for the updates may be derived as follows.

Specifically, the updates for $\{\hat{\upsilon}_i, \hat{\gamma}_i\}$ remain the same as in \eqref{upga}. However, the update for $\hat{\beta}$ takes the form:
\begin{align}\label{be2}
\hat{\beta} &= \frac{\omega}{\sum_{\tau=1}^t\sum_{i\in\Omega_\tau} \nu_{i\tau}}\\
\shortintertext{where,}
\nu_{i\tau} :=& y_{i\tau}^2 - 2(y_{i\tau}-\mu_e^{i\tau})(\mub^\A_i)^T\mub^\B_\tau - 2y_{i\tau}\mu_e^{i\tau} \nonumber\\
&+ (\mu_e^{i\tau})^2  + \Xi_e^{i\tau} + \tr{\Sig^\A_{i}\Sig^{\B}_{\tau,\tau}}.
\end{align}  
Further, the parameters $\mu_e^{i\tau}$ and $\Xi_e^{i\tau}$ are updated as
\begin{subequations}\label{eup}
	\begin{align}
	\Xi_e^{i\tau} &= \frac{1}{\hat{\beta} + (\mu_e^{i\tau})^2  + \Xi_e^{i\tau}} \label{36a}\\
	\mu_e^{i\tau} &= \hat{\beta} \Xi_e^{i\tau} (y_{i\tau} - (\mub^\A_i)^T\mub^\B_\tau). \label{36b}
	\end{align}
\end{subequations}

Proceeding similarly, the updates for $\{\mub^\J_i\}$, $\{\Xib^\J_i\}$, and $\{\Xib^\A_i\}$ remain the same as in \eqref{ja}, while the updates for $\{\mub^\A_i\}$ become:
\begin{align}\label{a2}
\mub^\A_i &= \hat{\beta}\Xib^\A_i\sum_{\tau \in\Omega'_i} \mub^\B_\tau (y_{i\tau}-\mu_e^{i\tau}) .
\end{align}
Finally, the updates for $\Xib^{\B}$ remain the same but the updates of $\mub^\B$ change. Specifically, the low complexity updates via LDL-decomposition remain mostly the same, except for the modified definition of $\v_\tau$ in \eqref{vbt} which now looks like
\begin{align}
\v_\tau = \hat{\beta} \sum_{i \in \Omega_\tau} (y_{i\tau} - \mu_e^{i\tau}).
\end{align}
The full robust subspace filtering algorithm is summarized in Algorithm \ref{alg:ALG2}. The predictions for $y_{i\tau}$ for $i\notin \Omega_\tau$ and for $\tau \geq t+1$ are obtained as in \eqref{pred1} and \eqref{pred2}, respectively. 
\begin{algorithm}
	
	Initialize $  \al,\gam,\bt,\v$, $sub=1,\, \Omega_\tau, \,\Omega'_i,\Xib^\A,\mub^\A, \Xib^\B,\mub^\B,\Xib^\J_{diag},\mub^\J \Lam_1,\mu_1, $
	
	$\hat{\mathbf{Y}}=\mub^\A(\mub^\B)^T$
	
	\While{$Y_{conv}< 10^{-5}$}{
		$\mathbf{Y_{old}}= \hat{\mathbf{Y}}$
		
		$\Gam=diag(\gam)$
		
		\uIf{$sub==1 $}
		{	
			$\text{Update using} \, \eqref{rts}$

			$sub=2$

			$ \text{Update using} \, \eqref{10a}, \,\eqref{sigma_up1},\,\eqref{15a},\,\eqref{15b}\,\forall \,\, 1\leq i \leq r$

		}
		
		\ElseIf{$sub==2 $} 
		{
			
			$\text{Update using} \, \eqref{sigma_up3},\eqref{15c}, \eqref{15d}, \eqref{10b} \, \,\forall \, 1\leq i \leq m$
			$sub=3$

		}
		
		\Else
		{
			$\text{Update using} \, \eqref{36a},\, \eqref{36b} \,\, \forall \,\, 1\leq i \leq m, \,\, \forall \,\, 1\leq \tau \leq t$

			$sub=1$
		}
		$\hat{\mathbf{Y}}=\mub^\A(\mub^\B)^T$

		$\text{Update using} \, \eqref{be2}$

		$Y_{conv}=\frac{\norm{\mathbf{Y}-\mathbf{Y_{old}}}_F}{\norm{\mathbf{Y_{old}}}_F}$
	}
	\Return($\hat{\mathbf{Y}},\Xib^\A,\mub^\A, \Xib^\B,\mub^\B,\Xib^\J_{diag},\mub^\J $)
	\caption{Robust Variational Bayesian Subspace Filtering}
	\label{alg:ALG2}
\end{algorithm}

%% file: result_conc_revised_pb_edit.tex
\section{Results}
\label{results}
We now detail the simulation results that evaluate the performance of the proposed VBSF method on variety of datasets to solve the: 
\begin{enumerate}
    \item Traffic Estimation and Prediction Problem
    \item Electricity Load Estimation and Prediction Problem
\end{enumerate}
\subsection{Datasets}
\begin{figure}
	\centering
	\includegraphics[width=0.7\linewidth]{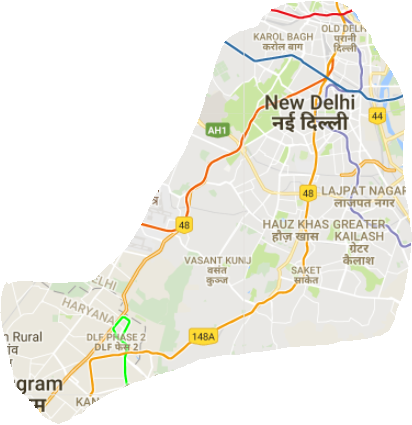}
	\caption{Region where traffic data is collected}
	\label{gmap}
\end{figure}
\begin{figure}
	\centering
	\includegraphics[width=0.7\linewidth]{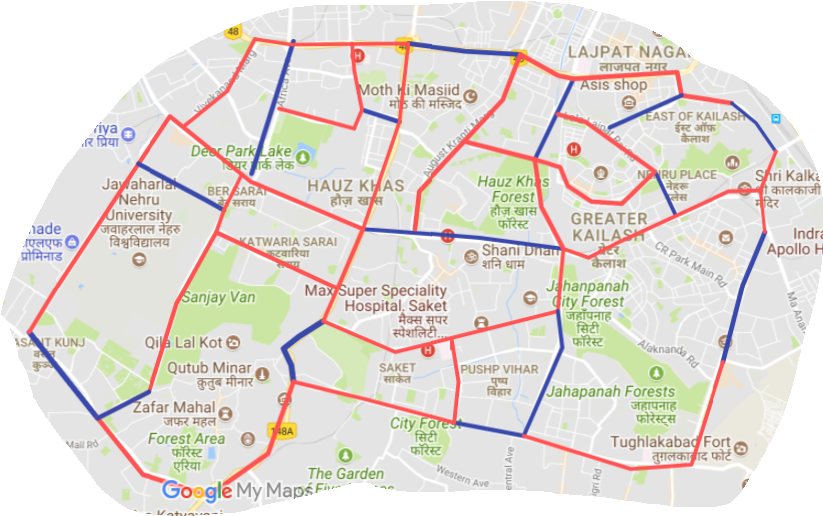}
	\caption{Map with red as missing and blue as known traffic entries}
	\label{gmap2}
\end{figure}
\begin{itemize}
	\item Traffic:  for traffic estimation and prediction, we use the partial road network of the city of New Delhi with an area of 200 square kms consisting of $m=519$ edges (shown in Fig. \ref{gmap}). The road network can be modeled using a directed graph where each edge represents a road segment and nodes represent intersections. We collect the traffic data in the form of average speed of vehicles on a particular segment using the Google map APIs for nearly 3 months across 519 edges. Taking advantage of the slow varying nature of the speed in the network edges, we sample the traffic data at the rate of one sample every $t_s =15$ minutes. Note that our algorithm is agnostic of the sampling rate and would  work for higher sampling rates as well. Unlike the complete data available from the API, real-world data may have missing entries. For instance, over the smaller area shown in Fig. \ref{gmap2}, speed measurements may be available on the blue edges but not on the red ones. Finally, we evaluate our algorithm for the twin tasks of real time traffic estimation as well as future traffic prediction. We further evaluate our algorithm for robust traffic estimation , i.e., when we the traffic data is corrupted by outliers.
	\item Electricity: similar to the traffic estimation and prediction task, we evaluate the VBSF algorithm on the electricity dataset \cite{electrictydata}, also used in \cite{paperarnew} to evaluate the online matrix factorisation method. The data contains the hourly power consumption of 370 consumers, sampled every 15 min. The data is recorded from Jan. 1, 2012 to Jan. 1, 2015. Finally, we compare the VBSF method with various methods including the ones proposed and compared in \cite{paperarnew}.
\end{itemize}
 In order to evaluate the VBSF algorithm, an incomplete data set is created by randomly sampling a fraction $p$ of the measurements. In our evaluations we consider three different cases with  75\%, 50\%, and 25\% of missing data.  We select previous $h$ = 30 time intervals for traffic and, the previous $h$ = 40 time intervals for electricity dataset. We compare our algorithm with other methods that potentially solve the current traffic estimation problem in the missing data scenario. The algorithms are 
 \begin{itemize}
\item Low rank tensor completion (LRTC) \cite{liu2013tensor}.
\item Grassmannian Rank-One Update Subspace Estimation (GROUSE) \cite{balzano2010online}.
\item Historic mean, which  is simply the mean of edge speed values at a given time instance calculated using the historic data. 
\end{itemize}
For the robust VBSF, we compare our algorithm with corresponding robust matrix completion frameworks. 
\begin{itemize}
    \item Robust PCA via Outlier Pursuit (OP-RPCA) \cite{oprca}.
    \item Robust Online Subspace Estimation and Tracking Algorithm (ROSETA) \cite{roseta}.
    \item  Grassmannian Robust Adaptive Subspace Tracking Algorithm (GRASTA) \cite{grasta}.
\end{itemize}
 Further, for the electricity load prediction problem, we compare our algorithm with the results of \cite{paperarnew} and the Collaborative Kalman Filter (CKF) \cite{paperarnew}.
\subsection{Traffic Estimation and Prediction Problem}
\subsubsection{Performance Index}
To measure the effectiveness of our algorithm and for the comparison with other relevant algorithms, we use mean relative error (MRE) as the performance index for the traffic data. For any time instance $\tau$, the MRE denoted by $\text{MRE}_\tau$ is defined as:
\begin{equation}
\text{MRE}_\tau= \frac{1}{z}\sum_{k=1}^z \frac {\parallel \hat{\y}_{\tau,k}-\y_{\tau,k}\parallel_{2}}{\parallel \y_{\tau,k}\parallel_2}.
\end{equation}
where $\y_{\tau,k}$ and $\hat{\y}_{\tau,k}$ are the ground truth and estimated data for $k^{th}$ day and $\tau^{th}$ time instance. Since the value for the known data (sampled entries) may be modified post estimation, we compute the MRE over the whole column for a given time instance. For calculating the overall accuracy of prediction for a day, we calculate MRE averged over $z$ days. The value of $z$ is taken as 50 for weekdays and 10 for the weekends.
\begin{figure*}[ht]
	\centering
	\includegraphics[width=1\linewidth]{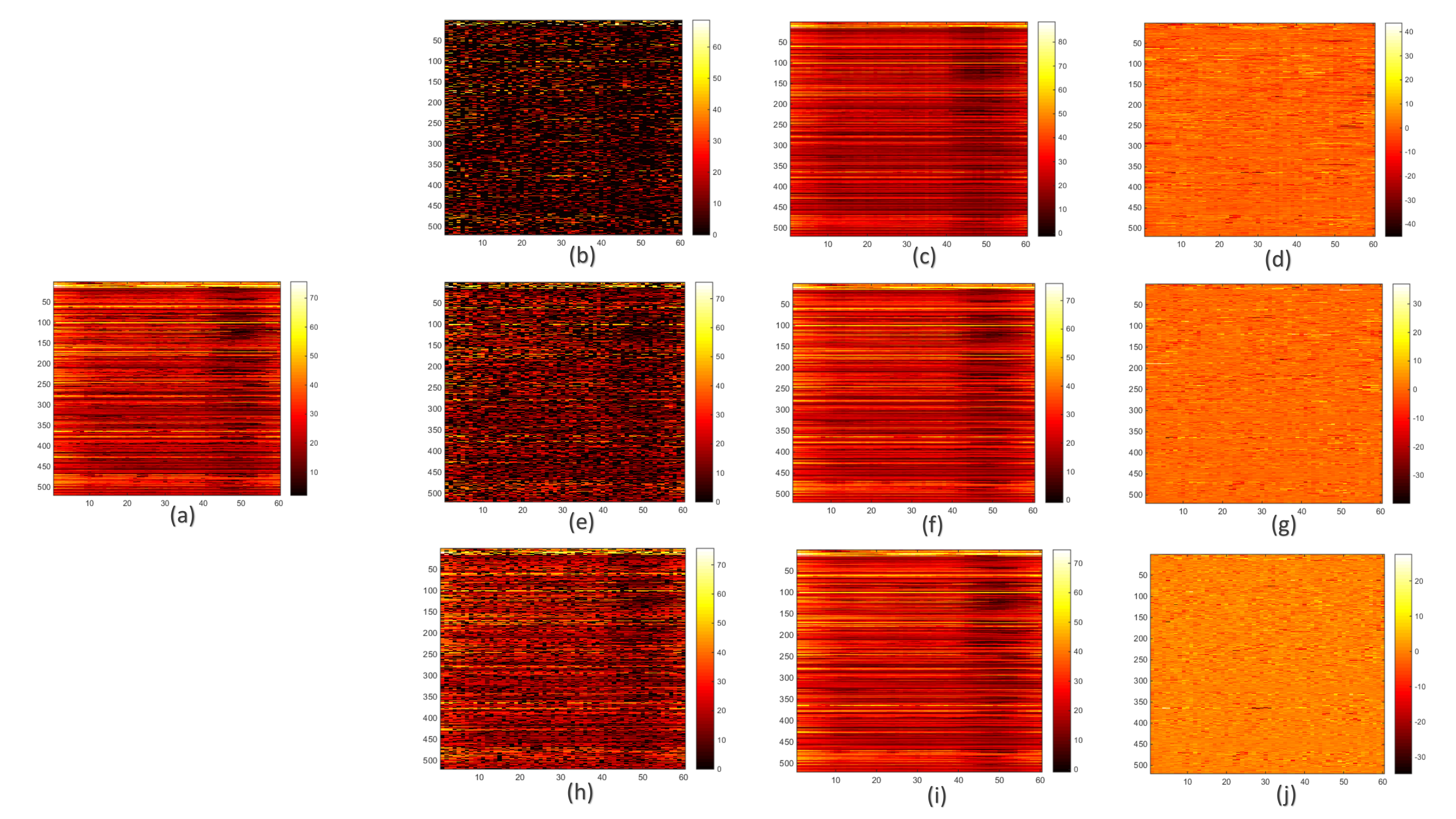}
	\caption{Estimation of traffic data for different percentage of missing entries}{(a) Actual Traffic data , (b) Traffic data with 25\% entries, (c) Estimated Traffic with 25\% known data, (d) Residual error for estimation with 25\% data , (e) Traffic data with 50\% entries , (f) Estimated Traffic with 50\% known data, (g) Residual error for estimation with 50\% data, (h) Traffic data with 75\% entries , (i) Estimated Traffic with 75\% known data, \\(j) Residual error for estimation with 75\% data }
	\label{fig:fplot}
\end{figure*}
\subsubsection{Online Real Time Traffic Estimation}
We now discuss simulation results for the current traffic estimation based on the current and past missing data using the VBSF algorithm. For a typical day, Fig. \ref{fig:fplot}a shows the heatmap of the actual traffic data. The $x$-axis of each  heatmap represents time instances while the $y$-axis represents the edges. Each pixel of a heatmap indicates the speed, where higher speed is represented by a lighter colour. Figures \ref{fig:fplot}b,  \ref{fig:fplot}e and \ref{fig:fplot}h are heatmaps with missing entries of varying degrees. The corresponding completed matrices using VBSF algorithm are shown in Figs. \ref{fig:fplot}c, \ref{fig:fplot}f, and \ref{fig:fplot}i.   Since the proposed VBSF is an online method that completes one column at a time given the incomplete data from previous columns, the corresponding heatmaps are also generated in an online fashion. In other words, in spirit of the online methodology, window of $h+1$ incomplete columns are used to complete the last column followed by moving the window by one column.  Finally, all the completed columns form a matrix represented in these heatmaps.  Unsurprisingly, the heatmaps show that the performance of VBSF improves as the size of missing data decreases.  \par
\begin{figure*}
	\centering
	\begin{subfigure}[b]{1\textwidth}	
		\includegraphics[width=1\linewidth]{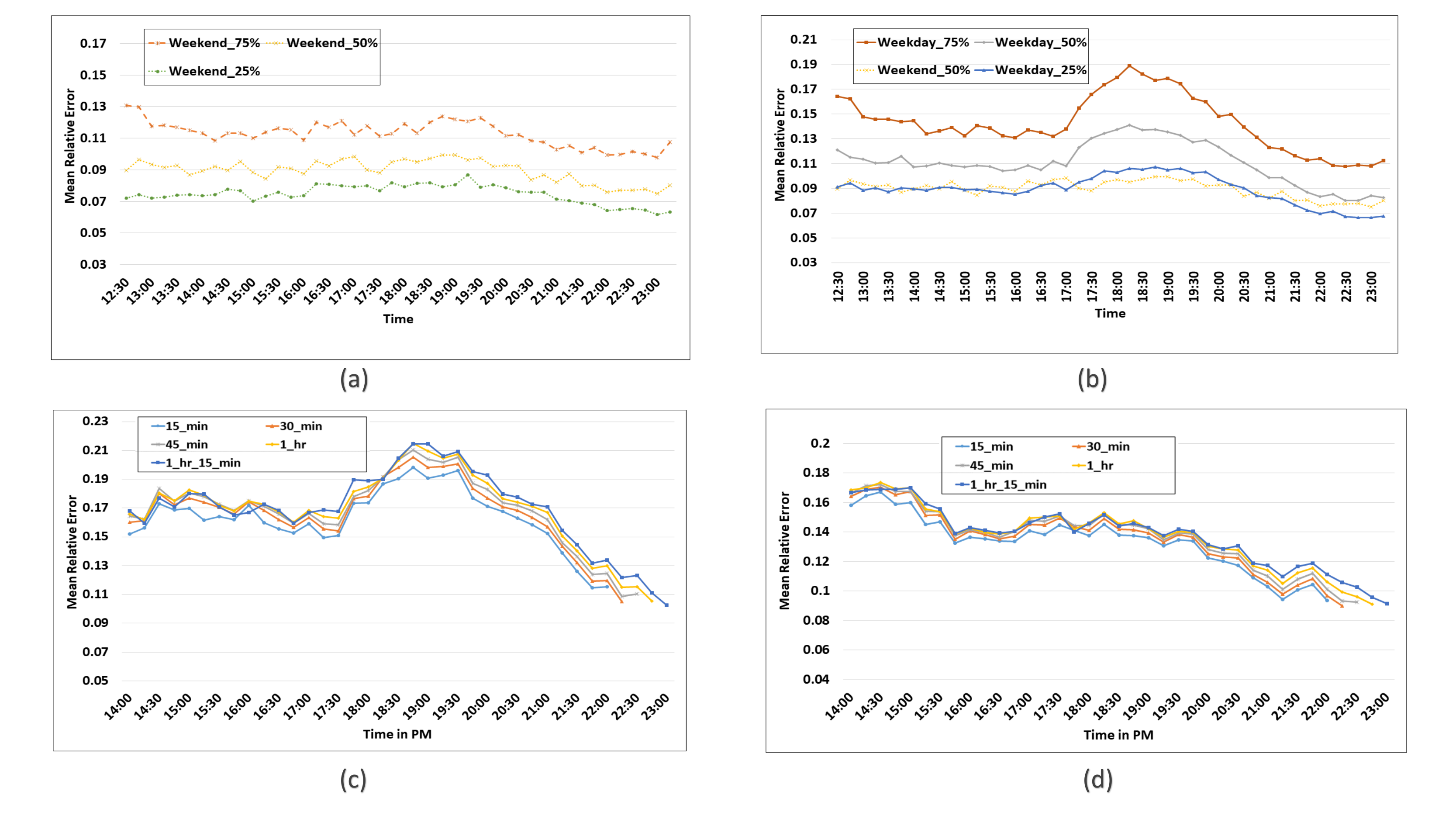}	
	\end{subfigure}%
	
	\begin{subfigure}[b]{1\textwidth}
		\includegraphics[width=1\linewidth]{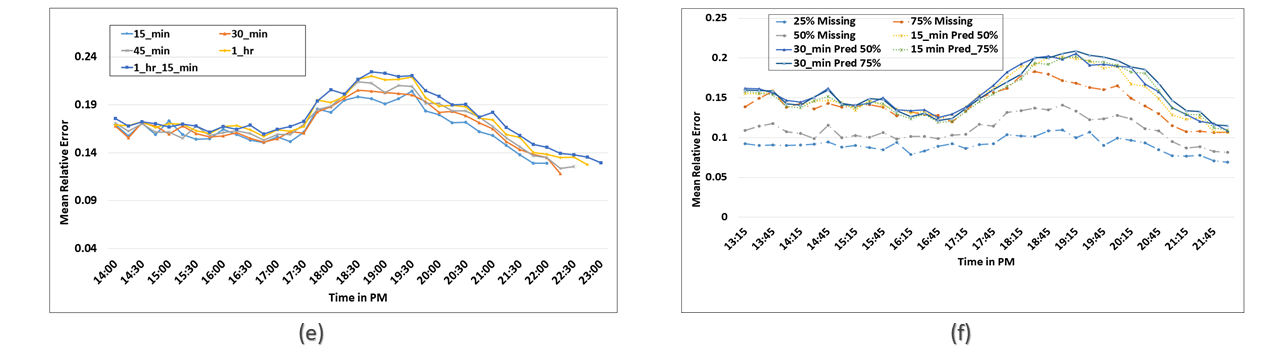}
		
	\end{subfigure}%
	\caption{Real time Traffic Estimation  and Prediction for different missing entries}{(a) Real time traffic estimation for different missing entries (Weekend), (b) Weekday Prediction 50\% missing entries (Weekday), (c) Weekday Prediction 50\% missing entries, (d) Weekend Prediction 50\% missing entries, (e) Weekday Prediction 75\% missing entries, (f) Overall Prediction}	
	\label{fig:fig4}	
\end{figure*}

  
The MRE values for real time traffic estimation using VBSF for weekends is shown in Fig. \ref{fig:fig4}a and for weekdays in Fig. \ref{fig:fig4}b. It is observed that the prediction error is higher during the peak traffic time (in the evening) vis-a-vis non-peak time intervals. This may be due to a greater variance in traffic during the peak time intervals. However, the difference between the MRE values for 50\% and 25\% missing data case is only about 0.15 in the worst case. Equivalently, the average error of estimation of speed is only around 2 km/hr during the peak-time when the average speed is 15 km/hr even with 75\% missing data.  Similarly, for non-peak hours, even though the observed speed are higher (around 30-40 km/hr), the MRE values for $p=50\%$ and $p=25\%$  is around 0.1, which in other words indicate an average error of 3-4 km/hr in the estimation of speed. \par 
The  performance  of  the  proposed  VBSF  algorithm  is compared   with   that   of  (LRTC) \cite{liu2013tensor}, (GROUSE) \cite{balzano2010online}, and  the historic mean. We used a grid search based approach for rank initialization in GROUSE and choose the rank that gives the least error.  Table \ref{tab1:table1} presents the overall results.  Further, Figs. \ref{fig:fig5}a and \ref{fig:fig5}b show the comparison of our algorithm for different percentage of missing traffic data. It is observed that for low missing rate of traffic data (25\%), the LRTC (low rank tensor completion) \cite{liu2013tensor} and VBSF obtain similar performance. But as the missing data increases,  VBSF outperforms the LRTC method.  Also, for all the cases, VBSF performs better than GROUSE.  This difference in performance can be attributed to the fact that the VBSF framework captures the temporal dependencies as well as the latent factors in the traffic matrix better than other methods. In terms of running time, VBSF is faster than LRTC and is comparable to GROUSE as shown in Table \ref{tab1:table12}. 

\begin{table}[ht!]
	
	\begin{center}

		\begin{tabular}{llll}
			\hline 
			&$p=0.25$ & $p=0.50$  &$p=0.75$\\
			&MRE&MRE&MRE\\
			\hline
			VBSF & 0.1439 &0.11277 &0.09336\\
			GROUSE & 0.372 & 0.3446& 0.3085\\
			LRTC & 0.1921 & 0.1418&0.09578\\
			Mean &0.2083&0.2083&0.2083\\
		\end{tabular}

	\end{center}
	\caption{Performance comparison for real time traffic estimation}
	\label{tab1:table1}
\end{table}

\begin{table}[ht!]
	
	\begin{center}
		\begin{tabular}{llll}
			\hline 
			&$p=0.25$ &$p=0.50$ &$p=0.75$\\
			&time($sec$)&time($sec$)&time($sec$)\\
			\hline
			VBSF &  0.7001&0.8685&0.9675\\
			GROUSE &0.7935&0.85324&0.923960\\
			LRTC &2.92&4.32&6.23\\
			
		\end{tabular}

	\end{center}
	\caption{Comparison of running time for different algorithms$^1$}
	\label{tab1:table12}
\end{table}
\footnotetext[1]{Experiments are conducted to evaluate average running time per column on Matlab using PC: Intel i5-6200U CPU 2.4 GHz. }
\subsubsection{Future Traffic Prediction Problem}
We also test the  VBSF algorithm for speed prediction during the future time intervals assuming randomly sampled data from the current and previous time intervals. We predict traffic data up to 5 sampling intervals, that is, 15 to 75 minutes in future.  We test our algorithm for 50\% and 75\% of the missing entries in the traffic data. The MRE plots for traffic prediction are shown in Figs. \ref{fig:fig4}c, \ref{fig:fig4}d, and \ref{fig:fig4}e.  The MRE error difference for 50\% and 75\% missing data is not significant. Similar to observations from the current traffic estimation simulations,  it is seen that the error increases from 5:30 to 8:00 pm. As one would expect, the prediction accuracy decreases as we predict further in future.  Interestingly, it is observed that the MRE for real-time traffic estimation with 75\% missing entries case and for future prediction with 50\% missing entries are comparable as can be seen in Fig. \ref{fig:fig4}f. \par
\begin{figure*}
	\centering
	\includegraphics[width=1\linewidth]{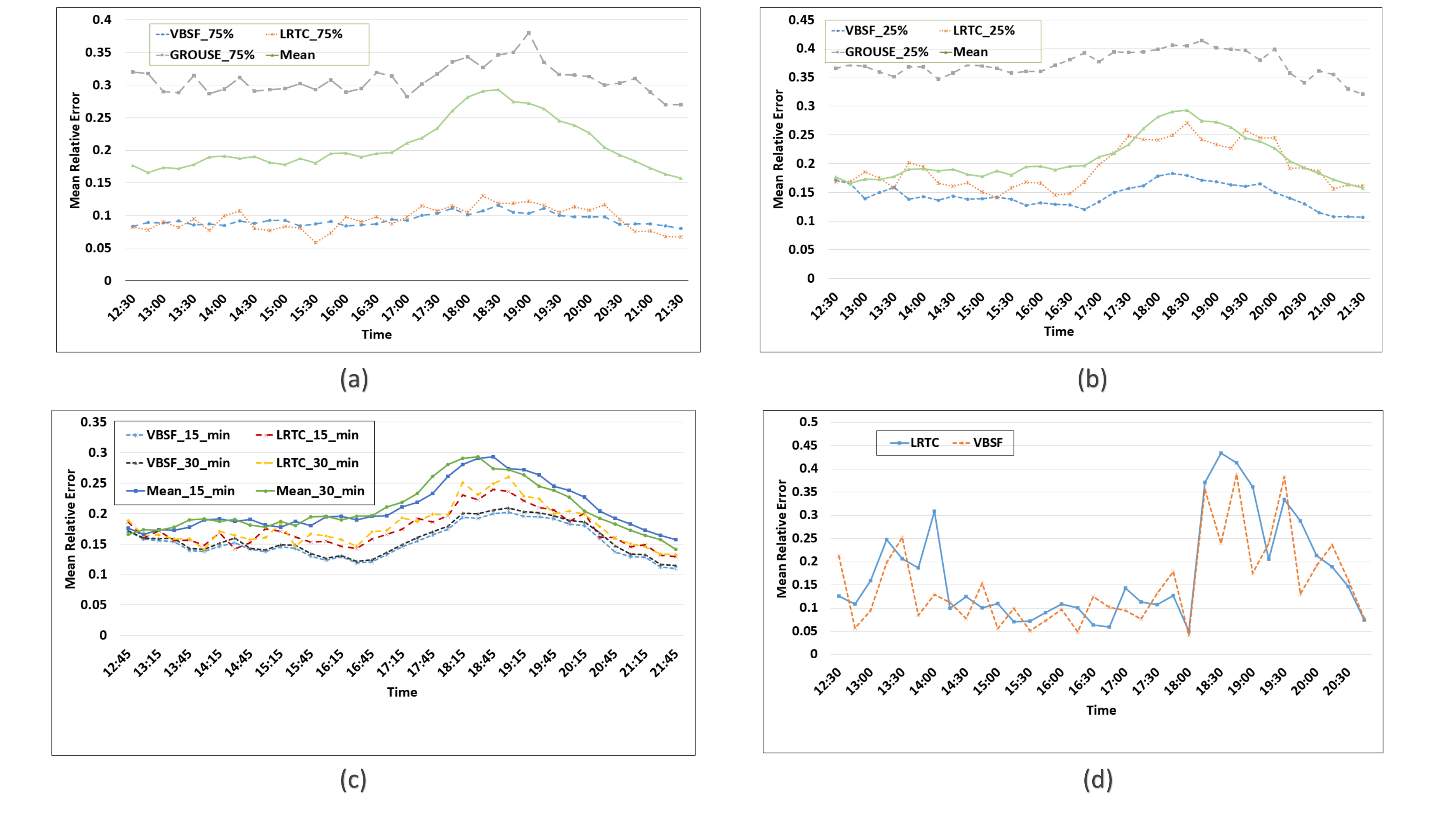}
	\caption{Comparison between VBSF and Low rank Tensor Completion (LRTC) and Matrix Completion Algorithm (GROUSE)}{(a) Real Time Traffic Estimation for 25\% percentage of Missing Data, (b) Real time traffic estimation for 75\% percentage of missing data, (c) Traffic prediction for 50\%  of missing data, (d) Traffic prediction for outliers}
	\label{fig:fig5}
\end{figure*}

 
The performance of the proposed VBSF algorithm is compared with that of LRTC in Table \ref{tab:table2}.  The VBSF performs better than the LRTC as shown in Fig. \ref{fig:fig5}c. While predicting the speed for outlier edges (the edges which significantly deviate from their usual speed) VBSF performs better than LRTC as seen in Fig. \ref{fig:fig5}d. 
\begin{table}[ht!]
	\begin{center}

		\begin{tabular}{lll}
			\hline
			&$p=0.50$ & $p=0.50$  \\
			&$15\, mins $&$30\,mins$\\
			\hline
			VBSF & 0.15362 &0.17434 \\
			LRTC & 0.15843 & 0.1812\\
			Mean & 0.2082 & 0.2073\\
		\end{tabular}

	\end{center}
	\caption{Performance comparison for traffic prediction }
	\label{tab:table2}
\end{table}

\subsubsection{Robust Traffic Estimation}
The GPS data that is collected using probe vehicles may be corrupted by noise and may often contain outliers which need to be removed before further processing is performed.  To mitigate the performance degradation due to outliers, we employ the robust variational Bayesian subspace filtering (RVBSF) that models the presence of outliers in the data in the sparse outlier matrix ${\bf E}$ . To test the RVBSF algorithm, on a given day, we randomly sample a certain $p_o$ percentage of the already sampled traffic data $\y_{i,\tau}$ and replace these values with $o_{i,\tau}$ as follows: 
\begin{equation} \label{outlier}
{\bf o}_{i,\tau} = \max \left( {\y}_{i,\tau-1},{\y}_{i,\tau+1} \right) + c \, \mu_t.
\end{equation}
In other words, the outlier is created by adding a large value $c\,\mu_t$ to the maximum of $\y_{i,\tau-1}$ and $\y_{i,\tau+1}$. Here, $\mu_t$ is the mean of observed entries at time $t$ and c is a scaling parameter. The RVBSF algorithm is then applied to solve the real time traffic estimation problem.  The detected artificial outliers are those points residing in the matrix ${\bf E}$.  \par
\begin{figure*}
	\centering
	\includegraphics[width=1\linewidth]{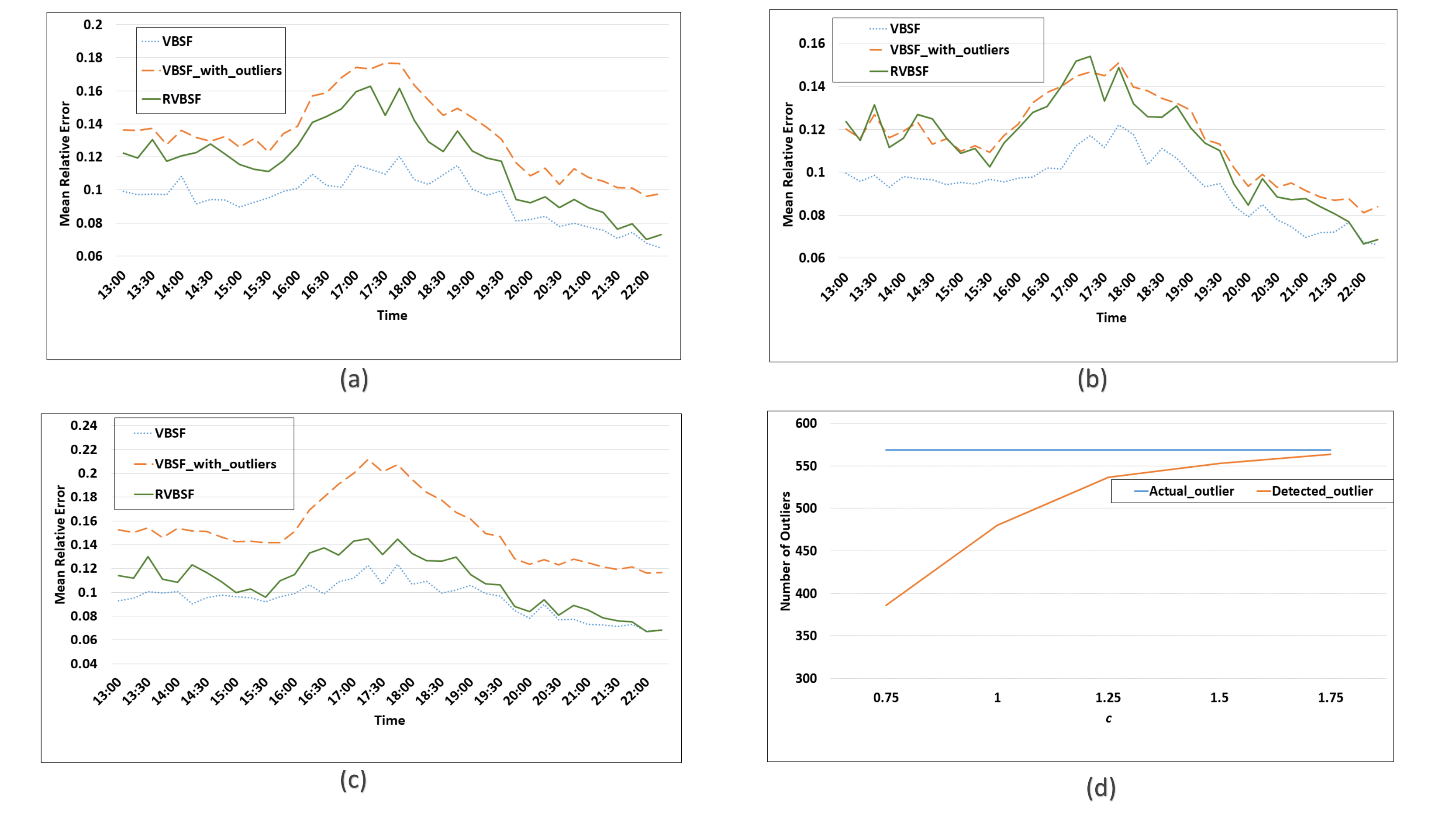}
	\caption{Robust Bayesian subspace filtering for traffic data }{}{(a) Comparison for VBSF and RVBSF with 5\% outliers and $c$ = 0.75, (b) Comparison for VBSF and RVBSF with 2\% outliers and $c$ = 0.75 (c) Comparison of VBSF and RVBSF for $c$ = 1.25, (d) Number of outliers detected for different outlier values}
	\label{fig:fig6}
\end{figure*}
The accuracy of outlier detection depends on the outlier value as shown in Fig. \ref{fig:fig6}d. The value of $c$ for simulations is chosen from the set $ [0.75, 1, 1.25, 1.5, 1.75]$.  We compare the robust VBSF (termed as RVBSF) with VBSF for two scenarios.  First,  when no outliers are added (VBSF), second,  when outliers are present in the data but only VBSF was used (VBSF\_with\_outliers). 
Table \ref{tab:table3} summarises the overall performance of the RVBSF algorithm. Understandably, RVBSF improves over VBSF when outliers are present, but is still worse than the MRE of VBSF for the case when no outliers were present. For 25\% missing entries, $p_o=5\%$ and $c=0.75$, the plots in Fig. \ref{fig:fig6}a illustrate the performance of the RVBSF algorithm.  Similarly for 75\% of missing entries, $p_o=2\%$ the results are shown in Fig. \ref{fig:fig6}b.  When $p_o = 5\%$ and $c= 0.75$, we observe that RVBSF detects outliers reasonably well vis-a-vis VBSF\_with\_outliers. Similar observation holds when outlier values increase as shown in Fig. \ref{fig:fig6}c and Fig. \ref{fig:fig6}d. \par 

\begin{table}[ht!]
	\begin{center}
		\begin{tabular}{llll}
			\hline 
			&$c=0.75$ & $c=0.75$ & $c=1.5$\\
			&$p_o=5$\%& $p_o=2$\% & $p_o=2$\% \\
			\hline
			VBSF & 0.09462 &0.09457 &0.09434\\
			VBSF\_outlier & 0.13406 & 0.11643& 0.15318\\
			RVBSF & 0.11741 & 0.1127&0.10912\\
		\end{tabular}
	\end{center}
	\caption{RVBSF: overall performance }
	\label{tab:table3}
\end{table}
The performance of the proposed RVBSF algorithm is compared with that of OP-RPCA\cite{oprca} GRASTA\cite{grasta} and ROSETA\cite{roseta} in Table \ref{tab:table4}. The RVBSF algorithm performs better than the subspace estimation and tracking algorithms. The difference in performance may be due to a better modeling of the temporal structure available in the data.  
\begin{table}[ht!]
\begin{center}
	\begin{tabular}{c c c c} 
		\hline
		
		&$c=0.75$ & $c=0.75$ & $c=1.5$\\
	&$p_o=5$\%& $p_o=2$\% & $p_o=2$\% \\
		\hline
		OP-RPCA& 0.2594 & 0.2298 & 	0.2165 \\ 
		\hline
		ROSETA& 0.1859 & 0.1819 & 	0.1723 \\ 
		\hline
		GRASTA& 0.1493 & 0.1507 & 	0.1492 \\ 
		\hline
		RVBSF & 0.11741& 0.1127& 0.10912\\
		\hline
	\end{tabular}
\end{center}
\caption{Performance Comparison for Robust Traffic Estimation }
\label{tab:table4}
\end{table}
A possible limitation of the suggested robust traffic estimation framework is following. While there may be outliers present due to an erroneous speed estimation, there might be cases when the  so called outlier value may actually be a real value. The current method may not be able to distinguish between such cases. Hence, a sudden drop in speed along an edge may be treated as an outlier and its possible impact on the traffic of nearby edges be be ignored by the model. 
\subsection{Electricity Load Prediction}
We now discuss the performance of the VBSF algorithm on the electricity load data set \cite{electrictydata}. Note that the electricity load data is also a time series data with the possibility of missing entries as well as temporal correlation between successive columns.
\subsubsection{Performance Index}
The performance of the VBSF method is compared with that of \cite{paperarnew}  using the metrics mean absolute error (MAE) and MRE, defined as: 
\begin{equation}
\text{MAE}= \frac{1}{z}\sum_{k=1}^z \frac {\parallel \hat{\y}_{k}-\y_{k}\parallel_{1}}{l(\y_{k})}
\end{equation}
\begin{equation}
\text{MRE}= \frac{1}{z}\sum_{k=1}^z \frac {\parallel \hat{\y}_{k}-\y_{k}\parallel_{2}}{\parallel \y_{k}\parallel_{2}}
\end{equation}
where $\y_{k}$ and $\hat{\y}_{k}$ are the ground truth and estimated data for $k^{th}$ column. We run the algorithm online on dates Jan. 1, 2012 to Jan. 1, 2015 resulting into 26,304 columns. In other words, the value of $z$ is 26,304 for our simulations.
\subsubsection{Online Electricity Load Estimation and Prediction}
	We run our algorithm for electricity data estimation and prediction. The results for real-time prediction are noted in table \ref{tab:ele1}. It is noted as the percentage of observed data $p$ increases, the real-time prediction accuracy improves. 
	
	\begin{table}[ht!]
		\begin{center}
			\begin{tabular}{llll}
				\hline 
				&$p=0.25$\%& $p=0.5$\% & $p=0.75$\% \\
				\hline
				
				MRE & 0.1789 &0.101 &0.0987\\
				MAE(kW) & 96.95 & 66.67& 53.95\\
				
			\end{tabular}
		\end{center}
		\caption{Electricity real time load prediction}
		\label{tab:ele1}
	\end{table}
	
\begin{figure}
	\centering
	\includegraphics[width=0.9\linewidth]{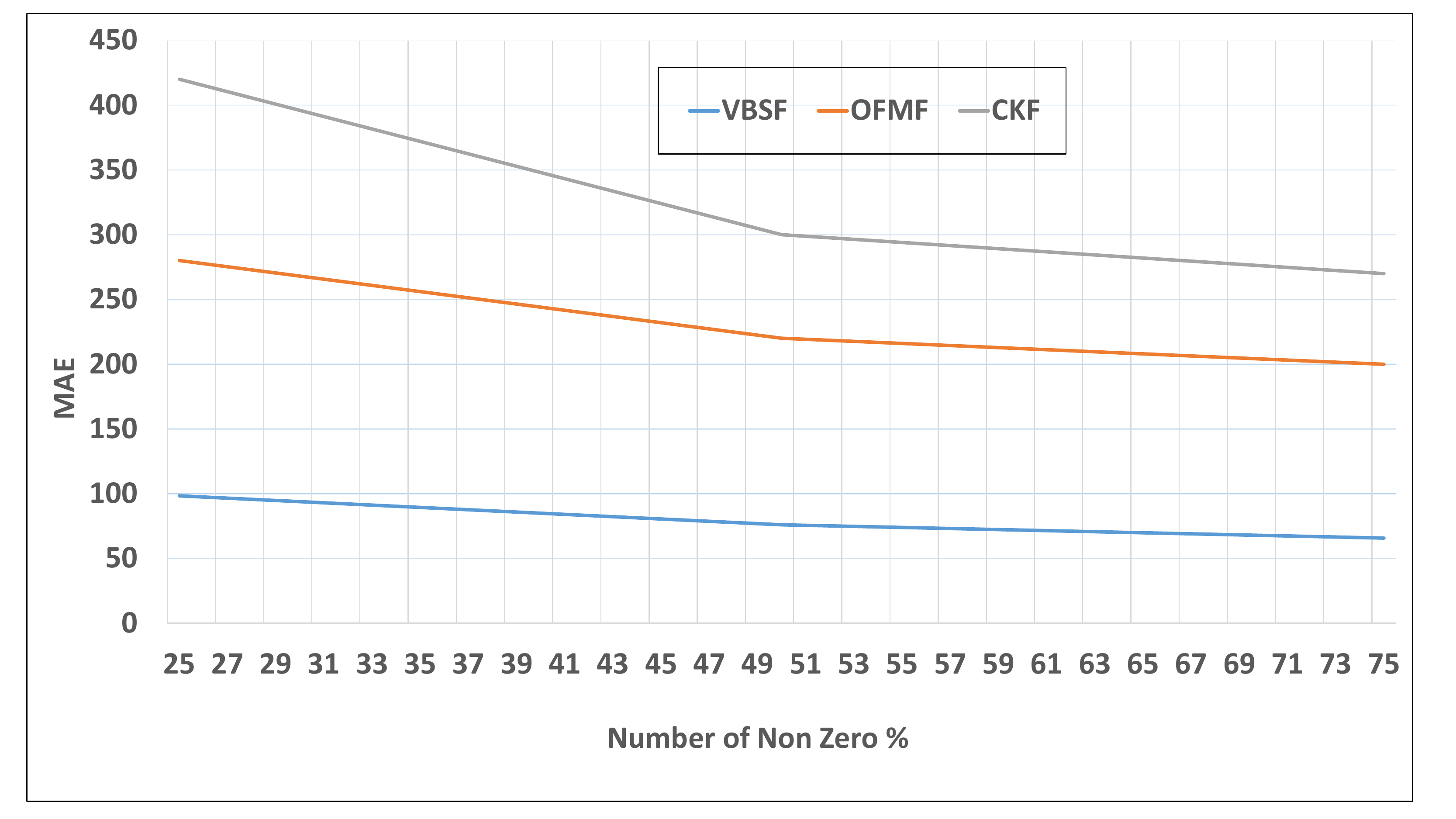}
	\caption{One-step ahead electricity prediction}
	\label{fig:elec}
\end{figure}

	Further, we predict the one-step ahead electricity load in Fig. \ref{fig:elec}. To analyze the performance of our algorithm we compare our results with OFMF and CKF \cite{paperarnew}. The one-step ahead prediction performance of OFMF and CKF are provided in \cite{paperarnew}. OFMF proposes a autoregressive model based optimization to predict the one-step ahead electricity load. We compare our three cases of $p$ with the results shown in OFMF. It can be seen that our algorithm performs better than the OFMF for electricity load dataset. 
	
\section{Conclusion}
\label{conclusion}
This paper considers sequentially arriving multivariate data that resides in a time-varying low-dimensional subspace. The temporal evolution of the underlying low-rank subspace is characterized via a state-space model and low-complexity variational Bayesian subspace filtering algorithms are proposed for matrix completion and outlier removal tasks. Simulation experiments quantify that the suggested model can be deployed to estimate the missing traffic data with a reasonable accuracy even with a fraction of random traffic measurements in the network. A similar result is observed on applying the VBSF algorithm on the twin tasks of imputation and prediction on the electricity data-set.  Extensive simulations on both the data sets demonstrate that the suggested model and the accompanying algorithms seem to capture the temporal evolution of the data well as compared to the current state-of-the-art matrix completion and the online subspace estimation algorithms.
